# Historical Evolution of Global Inequality in Carbon Emissions and Footprints versus Redistributive Scenarios

Gregor Semieniuk[*†] and Victor M. Yakovenko[‡]

Abstract:

Ambitious scenarios of carbon emission redistribution for mitigating climate change in line with the Paris Agreement and reaching the sustainable development goal of eradicating poverty have been proposed recently. They imply a strong reduction in carbon footprint inequality by 2030 that effectively halves the Gini coefficient to about 0.25. This paper examines feasibility of these scenarios by analyzing the historical evolution of both weighted international inequality in $CO_2$ emissions attributed territorially and global inequality in carbon footprints attributed to end consumers. For the latter, a new dataset is constructed that is more comprehensive than existing ones. In both cases, we find a decreasing trend in global inequality, partially attributed to the move of China from the lower to the middle part of the distribution, with footprints more unequal than territorial emissions. These results show that realization of the redistributive scenarios would require an unprecedented reduction in global inequality far below historical levels. Moreover, the territorial emissions data, available for more recent years up to 2017, show a saturation of the decreasing Gini coefficient at a level of 0.5. This observation confirms an earlier prediction based on maximal entropy reasoning that the Lorenz curve converges to the exponential distribution. This saturation further undermines feasibility of the redistributive scenarios, which are also hindered by structural tendencies that reinforce carbon footprint inequality under global capitalism. One way out of this conundrum is a fast decarbonization of the global energy supply in order to decrease global carbon emissions without relying crucially on carbon inequality reduction.

## 1. Introduction

The problem of mitigating climate change is increasingly linked with issues of inequality (Cantore and Padilla, 2010; Rao and Min, 2018; Zimm and Nakicenovic, 2019). At the international level, the Kyoto Protocol distributed responsibilities of climate change mitigation unequally across countries that also had unequal emissions and capacities to mitigate. But the United States did not ratify the treaty, because developing countries were not required to cut emissions (Schiermeier, 2012). More recently, this debate was revisited in unsuccessful negotiations in Copenhagen in 2009 about a follow-up treaty to Kyoto (Bailey, 2010). It ultimately led to countries pledging to do what their own governments thought was achievable, but not necessarily consistent with the global mitigation goal, the so-called Intended Nationally Determined Contributions under the 2015 Paris Agreement (Rogelj et al., 2016). At the national level, Jenkins (2014) warns that political economy constraints, in particular different abilities to pay and political resistance to regressive policies, may hinder policy implementation. Schor (2015) suggests that income redistribution could raise the chances for successful climate mitigation, and Boyce (2019) proposes a carbon dividend that redistributes the proceeds of a regressive carbon tax equitably among the population. In a more or less direct fashion, all of these socio-economic conflicts follow from the underlying problem that greenhouse emissions themselves are unequally distributed across countries and people.

[*] Political Economy Research Institute & Department of Economics, University of Massachusetts Amherst; & Department of Economics, SOAS University of London.
[†] Corresponding Author: gsemieniuk@econs.umass.edu
[‡] CMTC and JQI, Department of Physics, University of Maryland, College Park, MD.

It is now well known that, not only current per capita greenhouse gas emissions are vastly unequal across countries, but that cumulative emissions – which matter for greenhouse gas concentrations – are even more unequal. This leads directly to the argument that countries with higher per capita emissions should also be responsible for more per capita mitigation than poor ones (Pan et al., 2015). Increasingly, emissions inequality is also analyzed at the interpersonal level. According to one estimate, the highest 10% income earners are responsible for approximately 45% of global greenhouse emissions (Chancel and Piketty, 2015). Therefore, it is natural to ask how to take the inequality in carbon emissions into account when crafting mitigation strategies.

A prominent proposal for how to do this was advanced by Chakravarty et al. (2009) who suggested capping the level of per capita carbon dioxide ($CO_2$) emissions above a certain threshold. This cap would leave low emitters unaffected and shift the burden of mitigation to those with large emissions. Since emissions are highly positively correlated with income, this would also be equitable in the sense that the richest persons would have to deal with this problem. They showed that global $CO_2$ emissions could be reduced by 30% in 2030 if the billion highest emitters would cap their emissions at the level of their least polluting member. At the time of their writing, this reduction satisfied then prevailing climate change mitigation goals. We call it the "cap scenario" in the rest of the paper.

A related study by Hubacek et al. (2017) found that fulfilling the first of the United Nations' Sustainable Development Goals (UN SDGs), the eradication of poverty, would not only "lift" billions out of material poverty, but also raise their carbon footprint to the level of the non-poor due to the effect of higher purchasing power on emissions. Effectively, this would level emissions at the bottom of the distribution and put a floor under how little a person emits. Thus, we call it the "floor scenario" in the rest of the paper. While this paper was primarily concerned with the side effect on greenhouse gas emissions due to poverty reduction, it pointed out an important issue: that poverty reduction is intimately tied up with the distribution of emissions.

Surprisingly, while both of these proposals imply large reductions in emissions inequality, they did not attempt to quantify the change in carbon emissions or footprint inequality. Indeed, to the best of our knowledge, there is little research on the level and evolution of historical global carbon footprint inequality. Before reviewing the literature, it is useful to distinguish between different categories both of inequality and emissions. When analyzing inequality across countries, researchers usually refer to three different concepts of inequality (Milanovic, 2005). The first is international inequality, which takes one data point for each country, e.g. the average per capita carbon emissions. International inequality is typically considered in studies of convergence. The second type is weighted international inequality, where each observation is weighted proportionally to the country's population. Weighting serves to convey an impression of the global distribution and sometimes proxies for the third type, discussed next. The third type is global inequality, which does not look at countries as organizing units but instead at persons, or groups of persons, and their emissions distribution across the entire world. Here, belonging to a country's population is unimportant, although in practice estimates tend to use quantiles of country's populations and weigh them according to the country's size (Chancel & Piketty 2015). The cap and floor scenarios we just reviewed are falling into the global inequality category.

When talking about emissions, the most common measure is territorial emissions, that is the emissions produced in a country when burning fossil fuels. These are straightforward to calculate: Since there are good statistics about fossil fuel consumption, and the carbon content of these fuels is accurately known, it is a simple calculation to obtain territorial emissions. Then per capita emissions are just these total territorial emissions divided by the domestic population. We will refer to this measure as territorial emissions or simply $CO_2$ emissions. Carbon footprints, on the other hand, are calculated from "consumption-based" carbon accounting whereby all those emissions created in the production of all the final goods that are *consumed* in the territory are attributed to the country. The carbon footprint attributes the emissions from production to the final consumer, regardless of where they are emitted. National consumption-based and territorial emissions can vary considerably (Kander et al., 2015; Peters et al., 2011), and it is hotly debated which measure or a combination should be used to attribute responsibility (Tukker et al., 2020). For the purpose of interpersonal global inequality, it makes most sense, however, to consider the carbon footprint. Consumption expenditures of households are only indirectly related to territorial emissions (many of which are generated by the production of goods that are exported) but are directly linked to the carbon embodied in the goods purchased.

The literature can be classified according to these concepts. Most research has considered territorial emissions inequality, and here inequality between people within countries has been neglected relative to the attention given to international inequality (Islam and Winkel, 2017). Between countries, one focus is on convergence of country mean emissions, which finds convergence between groups of clubs, but no overall convergence (Aldy, 2006; Pettersson et al., 2014). These convergence studies do not typically quantify their results in terms of inequality indices. Another set of studies decomposes changes in (unweighted) international $CO_2$ emissions inequality (Grunewald et al., 2014; Heil and Wodon, 1997). Duro (2013) weighted and provided changes in the Gini coefficient, but did not report the actual inequality levels. Apart from these sets of studies, only Lawrence et al. (2013) and Zimm and Nakicenovic (2019) showed the evolution of weighted international inequality in territorial emissions since 1980 and 1990 respectively, which has been decreasing. Rao and Min (2018) used an estimated *increase* by 7 Gini points per decade in China's income in the past to argue this is a plausible upper limit for the rate of *decrease* in emissions inequality in general. Research on carbon footprints is scarcer. Wang and Zhou (2018) computed weighted international carbon footprint inequality but included only 40 regions. Closest to an evaluation of global carbon inequality may be Chancel and Piketty (2015), who compared how total footprint shares of various quantiles had changed and noted that between-country inequality had fallen, while within-country inequality had risen according to the Theil index. However, most of their evidence refers only to two years: 1998 and 2013 (the latter being extrapolated from 2008 income distribution and carbon footprint data), and the brunt of their analysis considers the 2013 snapshot. Overall, the literature on the level and evolution of global carbon footprint inequality is, at best, incomplete.

To improve our understanding of the ambitious scenarios for carbon emissions reduction, we analyze the historical evolution of inequality, both weighted international for territorial emissions and global for footprints. Section 2 quantifies the levels of carbon footprint inequality implicit in the "cap" and "floor" scenarios. In Section 3, we examine historical evolution of weighted international inequality in territorial emissions based on Energy Information Administration (EIA) data for 1980-2017 by extending previous estimates in Lawrence et al. (2013) and re-evaluating their prediction of an exponential distribution based on maximal entropy reasoning. In Section 4, we present the methodology for estimating global carbon footprints. We construct a new estimate based on a balanced panel of consumption expenditure data for twelve population quantiles from the Global Consumption and Income Project, GCIP (Lahoti et al., 2016) for most countries from 1995 to 2013, and a slightly smaller panel from 1970 to 1990, which excludes the former Soviet Union. Using a constant expenditure elasticity of footprints, we link consumption expenditure to the carbon footprint data from the multi-regional input-output database Eora for all years (Kanemoto et al., 2016; Lenzen et al., 2013). In Section 5, we compare our results with those by Chancel and Piketty (2015), as well as the estimates for the cap and floor scenarios. Based on the analysis of our results, in Section 6 we discuss challenges for the achievement of the scenarios drawing on the literature about global capitalism and the maximal entropy perspective.

We contribute to the research on carbon inequality by providing new evidence about the historical convergence of both weighted international carbon emissions inequality and carbon footprints toward an exponential distribution with a Gini of 0.5, as predicted by Lawrence et al. (2013). We also provide the first estimate of longer-term carbon footprint inequality evolution in the global economy and contrast it with climate change mitigation scenarios that often imply a quantifiable reduction in carbon inequality. While a more precise measure would require elasticities that vary by country, quantile, and over time (Grubler and Pachauri, 2009), we believe that using a constant elasticity gives us a good first approximation of the evolution of carbon inequality and countries' position within it over recent decades. An important take-away from our discussion is that policies seeking to harness reductions in inequality – whether at the top or bottom of the distribution – have to lean very heavily against persistent structural forces tending to high levels of inequality. Decarbonizing the energy supply can alleviate some of the need to grapple with carbon inequality for climate change mitigation, as it reduces emissions across the board.

## 2. Inequality in the cap and floor scenarios of carbon footprints

To start out, we illustrate the ability to quantify the extent of inequality in the cap and floor scenarios of future carbon footprints. In the cap scenario by Chakravarty et al. (2009), if the top billion emitters were to reduce their carbon footprint to those of the next billion by 2030, $CO_2$ emissions would be lowered in 2030 by 1/3 from 43 Gigatonnes in a business-as-usual scenario to 30 Gt, thereby reaching the climate change mitigation targets for the whole world estimated at the time of their writing.[1] This cap on top emitters' footprints implies eradicating inequality at the top of the emissions distribution, while leaving poorer emitters unaffected. More recently, Hubacek et al. (2017) instead calculated the impact on $CO_2$ emissions from putting a moderate floor under the income of everyone in the world (USD 2.97 per day or USD 1,084 per year) in the spirit of the first UN SDG. Abstracting from geographical and structural factors affecting carbon emissions, this puts a floor under carbon footprints and implies perfect equality at the bottom of the footprint distribution while leaving the top unaffected.

Although the inequality implications are obvious, neither proposal has attempted to quantify what impact on relative carbon footprint inequality their scenarios would have. In this paper we choose to quantify inequality via Lorenz curves. The Lorenz curve has considerable intuitive appeal and is widely used to analyze the evolution of inequality in energy consumption (Jacobson et al., 2005; Lawrence et al., 2013) and $CO_2$ emissions (Groot, 2010; Sager, 2019; Zimm and Nakicenovic, 2019). This is due to its graphical depiction of inequality as the divergence of the Lorenz curve from the perfect equality diagonal. Since the Gini coefficient is defined as twice the area between the perfect equality diagonal and the Lorenz curve, we use it as a numerical quantifier of choice alongside the Lorenz curve. For an introduction and formal definition of these measures see e.g. Cowell (2011) or Sen and Foster (1997).

---

[1] Chakravarty et al. actually considered territorial emissions, while the footprint datasets were still under construction at the time of their writing. Had footprint datasets been available, they could have been used instead of territorial emissions.

To apply these measures, we implement the two scenarios on a 2010 empirical estimate of the global distribution of carbon footprints that Hubacek et al. (2017) also provide. This distribution modifies the World Bank's Consumption Database to partition the world's population into 5 income brackets, the lowest comprising 15.5% of the population who earn less than USD 1.90 per day, then the 34.5% of global population earning between USD 1.9 and USD 2.97 a day and three more such quantiles representing the remaining 25%, 15% and 10% of the global population. Since these data also provide information on the basket of goods consumed, Hubacek et al. (2017) can link this to the demanded goods' carbon footprints supplied sector by sector from the Eora multi-regional input-output database, and so calculate the per capita carbon footprint for each quantile. With this global distribution at hand, the cap scenario simply scales down the top quantile's footprint to the level of the second one, and then both further until total emissions become two thirds of the initial estimate. The floor scenario scales up the bottom two quantiles' per capita footprints to the level of the third one, as a result of raising incomes assumed by Hubacek et al. It is also possible to combine the two scenarios. Then the bottom two quantiles are scaled up to that of the third, while the top two quantiles must scale down their carbon footprint even more in order to reduce total emissions back to those of the cap scenario. The cap+floor scenario can be thought of as simultaneously eradicating poverty and neutralizing the additional emissions.

Figure 1 illustrates these quantifications using Lorenz curves. The cap scenario (blue triangles) shows a straight line for the top two quantiles that include the highest 25% of emitters, i.e. equality at the top of the distribution. The floor scenario (red squares) instead shows proportional increases in population and share of carbon footprints, i.e. equality, for about the 75% lowest emitters. The combined scenario (green disks) is close to global equality (the straight diagonal line). These curves translate into Gini coefficients of 0.32 (cap), 0.27 (floor), and 0.07 (both). For a reference, we also plot the Lorenz curve of an exponential distribution discussed in Sec. 3, which corresponds to a Gini of 0.5. Historical evolution of the Lorenz curves for carbon emissions and footprints will be shown in Figures 2 and 4 discussed in the corresponding sections of the paper. It is clear that, under any proposed scenarios, carbon footprint inequality would be very low compared to the actual estimates of global carbon inequality, which is more unequal than an exponential distribution and has Gini coefficient above 0.5. A realization that these inequality reduction targets are implicitly present in the current goals by the governments of the world (UN SDGs and Paris Agreement) raises the question whether such proposals are realistic, given how distant they are from the empirical carbon Lorenz curves and Gini coefficients, discussed in more detail in the rest of the paper.

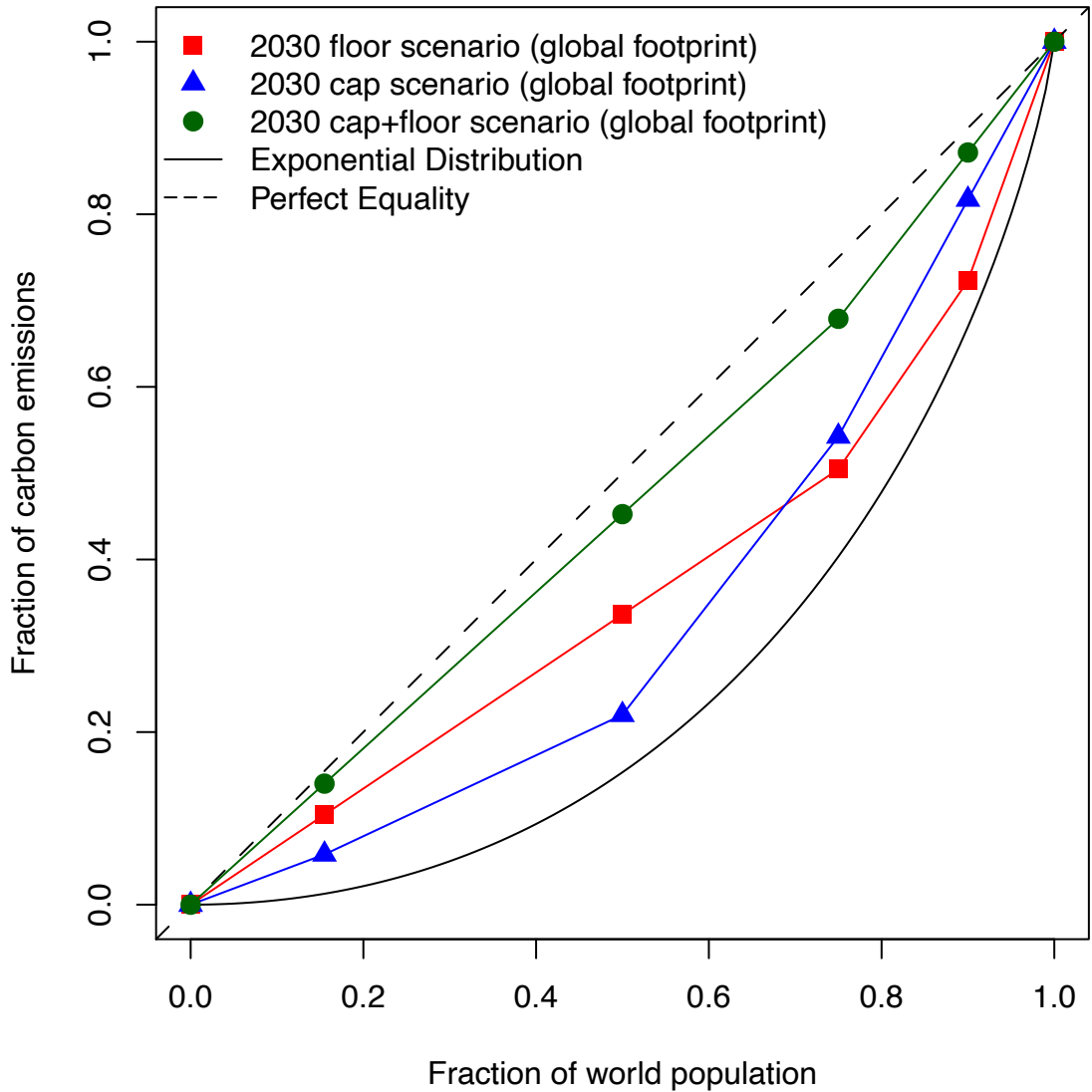

Figure 1: Lorenz plots for the global carbon footprint per capita estimated for five income groups for three scenarios that variously eradicate inequality among the bottom half (red squares), the top billion of the population (blue triangles), and both scenarios combined (green disks) relative to 2010 data. Perfect equality (dashed black line) and an exponential distribution (solid black line) are plotted for comparison. Data construction is discussed in Section 2.

## 3. Territorial inequality in $CO_2$ emissions and the maximal entropy interpretation

To assess the historical evolution of inequality, we first investigate the distribution of territorial carbon emissions, which is straightforwardly available. Lawrence et al. (2013) computed weighted international inequality, i.e. one observation per country, using the U.S. Energy Information Agency (EIA) data on population and territorial $CO_2$ emissions for 1980-2010. Since then, new data became available for subsequent years, so we show the updated Lorenz curves through 2017 in Figure 2. One can see a significant *decrease* of carbon emissions inequality from 1980 to 2010, the period studied by Lawrence et al. (2013), which they interpreted as evolution toward the exponential distribution corresponding to maximal entropy. The paper predicted that further evolution of the Lorenz curve and the decrease of inequality would stop once the exponential distribution is reached. Indeed, Figure 2 shows no significant change from 2010 to 2017, and both curves are close to the exponential distribution. Thus, the new IEA data for 2011-2017, which became available after publication of their paper, confirm the prediction by Lawrence et al. (2013) that the Lorenz curve evolution saturates at the exponential distribution.

Interestingly, the empirical data points for 2010 (not to be confused with the hypothetical "floor scenario" for 2030 discussed in Sec. 2) from Hubacek et al. (2017), obtained by a very different method, also approximately fall on the Lorenz curve corresponding to the exponential distribution, as shown by red stars in Figure 2. It is encouraging that the independent and methodologically different studies by Hubacek et al. (2017) and by Lawrence et al. (2013) produce qualitatively consistent Lorenz curves for 2010. While territorial emissions are not the best for examining global inequality, the EIA data analysis shown in Figure 2 serves as a useful benchmark for comparison with other measures, because of the wide coverage (available since 1980 and for the largest part of the world, especially after 1990) and the least amount of imputation (in contrast to Sec. 4).

Furthermore, Lawrence et al. (2013) also observed a decline of the global Gini coefficient for $CO_2$ emissions from 0.67 to 0.52 and predicted an S-shaped time evolution of the Gini coefficient, saturating at 0.5 when the exponential distribution is achieved. The new data points that became available from EIA for the years 2011-2017 after publication of their paper confirm this prediction. These new data points indicated by red squares in Figure 3 clearly show saturation at the 0.5 level. The theoretical prediction of this saturation is particularly striking, because the black circles in Figure 3, as published in Lawrence et al. (2013), do not give any indication of saturation yet and could be naively (but incorrectly) extrapolated along the straight dashed line indicating a linear decrease of inequality.

The remarkable alignment with, and convergence to, an exponential distribution suggests a general mechanism at work in keeping inequality above the 0.5 Gini level. Banerjee and Yakovenko (2010) and Lawrence et al. (2013) developed a novel approach to predicting global inequality in energy consumption and carbon emissions based on the maximal entropy principle. In this framework, a global pool of extracted fossil fuels is distributed by transnational corporations for consumption by human population. Although, hypothetically, fossil fuels can be partitioned equally among the human population (corresponding to a zero Gini, somewhat like in the cap+floor scenario discussed in Sec. 2), such a distribution is extremely improbable. In contrast, based on combinatorial arguments, the most probable distribution is the one that maximizes entropy of the partition, subject to the constraint on the total fossil fuel pool. Mathematically, this distribution is described by the exponential function, which was earlier discussed in the context of monetary and income inequality (Dragulescu and Yakovenko, 2001, 2000; Tao et al., 2019; Yakovenko, 2013; Yakovenko and Rosser, 2009). Conceptual aspects of entropy applications to economic and social problems are reviewed by Rosser (2016), and antagonism between entropic and anti-entropic social forces is discussed by Rosser (2020).

The partitioning model is particularly suitable for the study of territorial inequality in $CO_2$ emissions, because fossil fuels are readily transportable on global scale (especially coal and oil, but less so natural gas) and then are burned to produce $CO_2$. Lawrence et al. (2013) also applied a similar argument to global inequality in energy consumption, under assumption that most energy comes from fossil fuels. This assumption is applicable only approximately. A more recent study by Semieniuk and Weber (2019) found a significant difference between the analyses based on the EIA data in Lawrence et al. (2013) and other data sources using different methodologies for energy accounting.[2] Nevertheless, all of these data show an S-shaped time evolution of the global Gini coefficient for energy consumption (Semieniuk and Weber, 2019). This behavior is qualitatively similar to the global Gini coefficient for $CO_2$ emissions shown in Figure 3. On the basis of their model, Lawrence et al. (2013) made a policy proposal that a transition from fossil fuels to renewable energy would not only address the environmental problems but also reduce global inequality in energy consumption. Solar and wind energy can be locally produced and locally consumed, particularly on a small scale, thus making energy consumption more uniform around the world. This is in contrast to the extraction-and-repartitioning model for fossil fuels, where some oil-producing countries (e.g. Nigeria) consume very little of what is extracted from their land.

Other socio-economic variables also appear to follow distributions that can be well explained by maximal entropy subject to plausible constraints (Scharfenaker and dos Santos, 2015; Scharfenaker and Semieniuk, 2017). If constraints vary, so does the shape of the distribution (Shaikh, 2017). Therefore, maximal entropy appears to provide a useful description of inequality given the socio-economic constraints. We will revisit this argument again in the discussion in Section 6.

---

[2] A similar mechanism is likely to explain the lower 0.4 Gini estimate obtained by Zimm and Nakicenovic (2019) for all territorial GHG emissions as opposed to only $CO_2$, as the territorial emissions from non-$CO_2$ sources (significantly from agriculture) are likely to be less correlated with income and its inequality than $CO_2$.

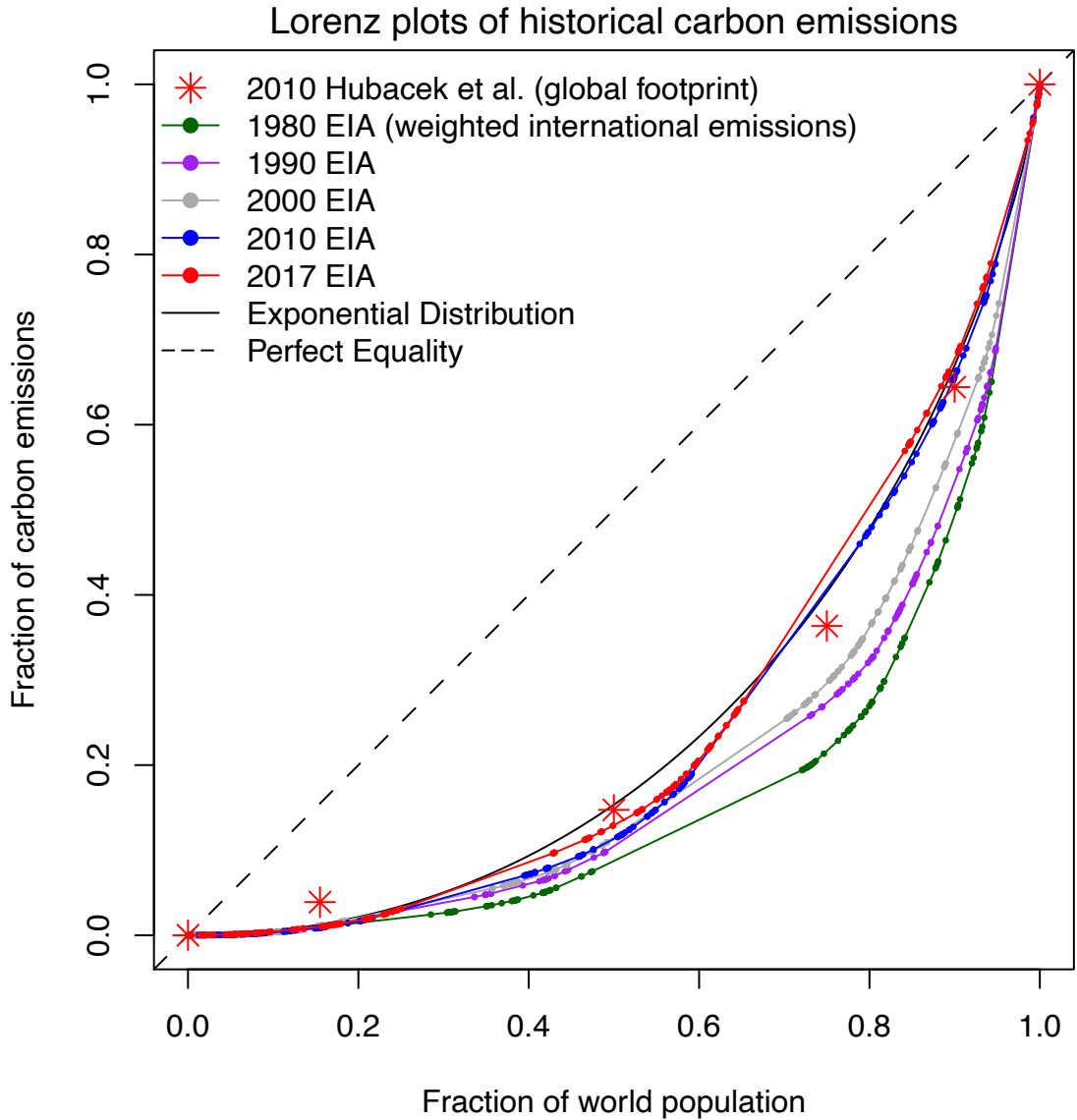

Figure 2: Historical evolution of the global Lorenz curve for selected years for weighted international per capita territorial carbon emissions based on EIA data. Each dot indicates a country's position. Red stars represent five empirical carbon footprint quantiles from Hubacek et al. (2017) for 2010 only.

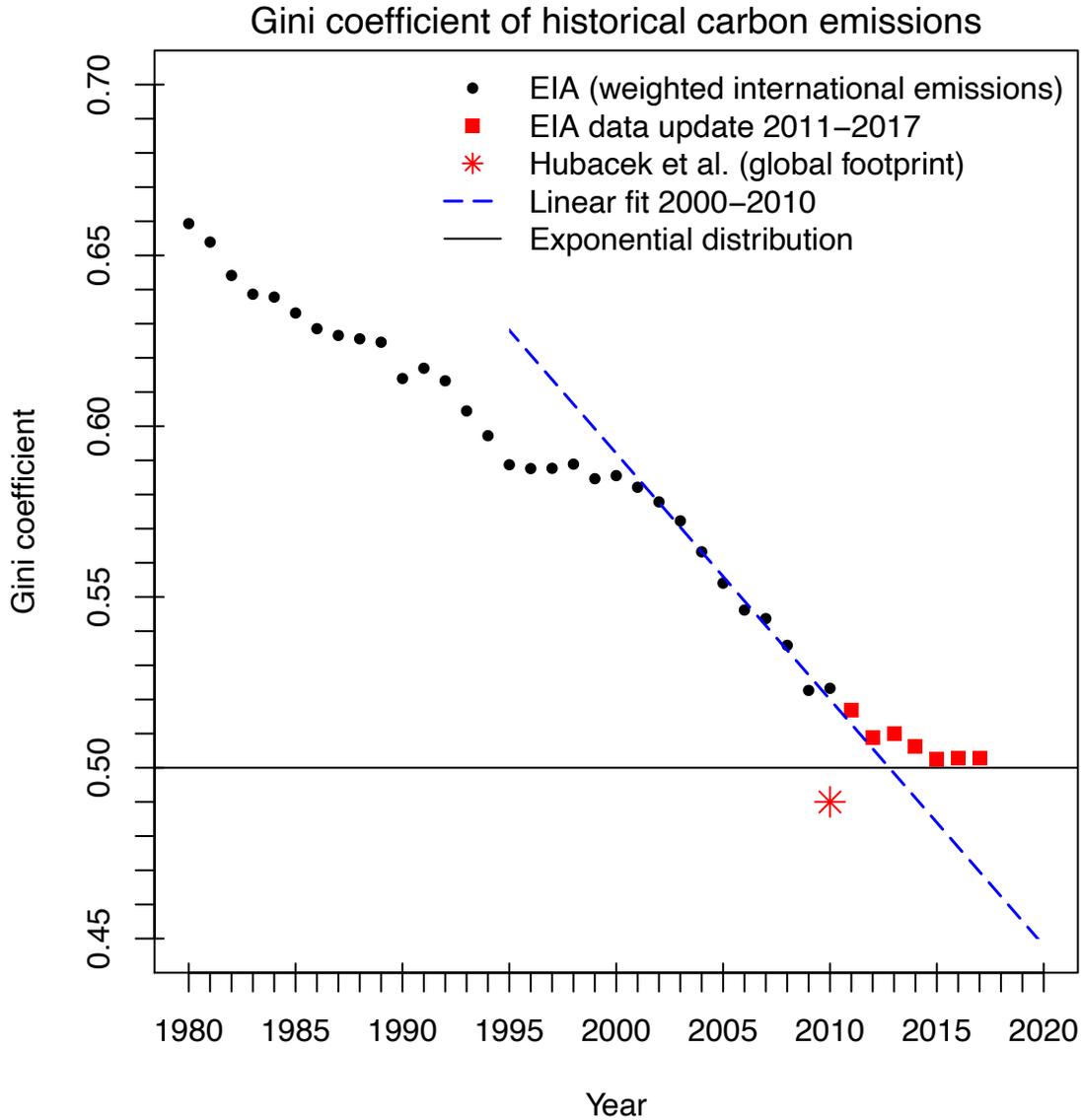

Figure 3: Historical evolution of the global Gini coefficient for the weighted international per capita territorial carbon emissions. Black disks show the data for 1980-2000 as published by Lawrence et al. (2013), whereas red squares show the new EIA data for 2011-2017. The blue dashed line is a linear extrapolation from 2000-2010. The horizontal line at Gini = 0.5 corresponds to a theoretical exponential distribution.

## 4. Methodology for calculation of global carbon footprints

To make further progress with global inequality analysis and to differentiate per capita emissions within a country, it is useful to look at income or expenditure surveys and attribute a certain share of the total national carbon footprint to each income or expenditure quantile. Here we relate aggregate national carbon footprints to national distributions of personal or household consumption expenditure by using a constant elasticity of carbon footprint with respect to expenditure, as is common in the literature (Chakravarty et al., 2009; Chancel and Piketty, 2015; Rao and Min, 2018). The result is a national distribution of carbon footprints. Repeating this for all countries approximates a global distribution.

While some authors use the discrete expenditure datapoints to estimate a footprint density across the range of personal expenditures, here we calculate one carbon footprint for each reported expenditure quantile without making further assumptions about the functional form. Following Chancel and Piketty (2015), country and quantile specific carbon emissions $C_{iq}$ are arrived at as

$$C_{iq} = C_{itot} \frac{f_{iq} y_{iq}^{\eta}}{\sum_q f_{iq} y_{iq}^{\eta}} \qquad (1)$$

where countries are indexed by i, quantiles by q, y is the mean expenditure in the quantile, $\eta$ is the elasticity of carbon footprint with respect to expenditure, and f is the share of the quantile in country i's population. $C_{itot}$ are country i's total emissions, so the second factor partitions these across quantiles.

For the new dataset constructed here, data on consumption expenditure and population are taken from the Global Consumption and Income Project (GCIP) by Lahoti et al. (2016), which reports consistent expenditure data on twelve consumption quantiles. These are the bottom nine deciles, the 90-95th percentiles, the 95-99th percentiles, and the top one percentile, for 162 countries in five year or smaller intervals from 1960 to 2013.[3] National carbon footprints are from Eora's carbon footprint database, which computes these annually for 192 countries from 1970 (Kanemoto et al., 2016). Eora is a multi-regional input-output database that allows tracing carbon emissions embodied in trade to their final consumption destination (Lenzen et al., 2013). The intersection of the countries in the GCIP expenditure database and Eora carbon footprint database is 150, comprising 95.8% of the global population in 2013. Prior to 1991, data on the Soviet Union is missing, thus reducing the share of population to 89.8% in 1970.

To compute quantile-wise emission footprints, equation (1) is applied to these GCIP-Eora data. An elasticity of 0.9 is chosen as the central elasticity estimate. Elasticities can vary between countries, expenditure quantiles, urban and rural dwellers and over time, therefore picking an average elasticity can only approximate the actual distribution. At the least, expenditure estimates are better correlated with emissions than income, as the latter can be saved and savings propensities differ strongly between income quantiles (Jonathan Fisher et al., 2018; Taylor et al., 2017). An elasticity of 0.9 is a conservative choice and appears to fall near the bottom of recently estimate of elasticities for large developing countries, so we are careful not to overstate inequality (Grubler and Pachauri, 2009; Wiedenhofer et al., 2017). 0.9 is also considered a plausible number for Rao and Min (2018) and used by Chancel and Piketty (2015) as their central estimate. While this approach is necessarily preliminary, it does serve to capture the approximate shape and evolution of $CO_2$ footprint inequality. Table 1 summarizes the data for this estimate. All Lorenz plots use this elasticity, but to check robustness, we also present Gini estimates for an elasticity of 0.7 (what Rao and Min (2018) consider a very low elasticity), and the mirror image of 1.1.

Table 1: Summary Stats GCIP-Eora, in metric tons $CO_2$ per capita for n countries times 12 quantiles (bottom 9 deciles, 90-95th vigintile, 95-99th quantile and 99-100th percentile for selected

---

[3] A subset of data is also available for 2015. However, there are significant gaps, so we restrict our analysis to the latest complete year 2013.

years, conditional on an expenditure elasticity of $CO_2$ emissions of 0.9. Median, mean and standard deviation (SD) are weighted by population.

| Year | n (countries) | Min | Median | Max | Mean | SD |
|---|---|---|---|---|---|---|
| 1970 | 133 | 0.008 | 0.838 | 271.762 | 3.863 | 66.807 |
| 1995 | 149 | 0.006 | 1.344 | 666.367 | 3.987 | 52.574 |
| 2013 | 150 | 0.011 | 2.200 | 639.397 | 4.820 | 62.568 |

We also draw on the greenhouse gas footprint dataset constructed by Chancel and Piketty (2015) for comparison with our results. One part of this dataset is the World Panel Income Database, WPID (Lakner and Milanovic, 2016) with 10 income decile data until 2008. Another part is national greenhouse gas (GHG) footprints for 1997/8 and 2007/8 from the Global Trade Analysis Project, GTAP (Andrew and Peters, 2013). Finally, Chancel and Piketty augment these datasets with their own World Top Income Database, WTID (Alvaredo et al., 2016) for 1998, 2003 and 2008, and extrapolate WPID and GTAP data to arrive at an additional datapoint in 2013. This last datapoint relies on strong assumptions about continuity in emission and income patterns, which must be taken with a grain of salt given the shock these variables sustained from the Great Recession.

In Section 2, we already introduced the data underlying cap and floor scenarios. For the cap scenario, Chakravarty et al. (2009) estimate 7 income quantiles for every country in their dataset based on various datasets from the World Bank: two deciles at bottom and top each, and three quintiles in between. These are combined with territorial emissions from the Energy Information Agency (EIA) for 2003 split across 7 quantiles using a constant elasticity of 1 and estimating a best fit theoretical distribution. To obtain a 2030 distribution, they take EIA projections for emissions and populations and assume the within-country relative income distributions to maintain their 2003 shape. In Section 2, we also explained Hubacek's dataset at length. They also arrive at their 2030 distribution by assuming that the relative income distribution and emissions intensity above the floor stay as in 2010. Lastly, there is the EIA territorial data itself, described in Section 3.

Table 2 compares the various datasets used in terms of their coverage. Note that the dataset used by Chancel and Piketty includes also non-$CO_2$ greenhouse gases.

Table 2: Comparison among datasets used

|  | GCIP-Eora (new) | Chancel & Piketty | Hubacek et al. | Chakravarty et al. | EIA (Lawrence et al.) |
|---|---|---|---|---|---|
| Emissions covered | $CO_2$ footprint | All GHGs footprint in $CO_2$ equivalents | $CO_2$ footprint | Territorial $CO_2$ | Territorial $CO_2$ |
| Period covered | 1970-2013 | 1998-2008 (extrapolated to 2013) | 2010, scenario for 2030 | 2003, scenario for 2030 | 1980-2016 |
| Countries covered in last year of data | 150* | 94 | 93+EU | 154** | 199 |

| Share of global population covered in last year of data | 95.8% | 86.7% | 89.6% | 99.5%** | >99.5% |
|---|---|---|---|---|---|
| Quantiles | 12 | 11 | 5*** | 7** | 1 |

*Prior to 1990, the countries covered are 133.
** 23 countries report only one average per capita $CO_2$ emissions observation.
***The size of quantiles is set at the global level.

## 5. Global inequality in carbon footprints

We proceed to our analysis of the evolution of historical global carbon footprint inequality by plotting the Lorenz curves for selected annual distributions from the newly constructed GCIP-Eora data, as described in Section 4. Figure 4 shows that Lorenz curves are far towards the bottom right corner, indicating a very high global inequality, higher than in the weighted international estimates shown in Figure 2. It is generally expected that the global carbon footprint inequality is higher than the weighted international territorial carbon emissions inequality. First, rich countries tend to have higher carbon footprints than territorial emissions, and second, global inequality takes into account inequality within countries, whereas the latter does not. Over time, the Lorenz curves in Figure 4 shift inward, so that each subsequent curve Lorenz-dominates the previous one (i.e. curves do not intersect). Thus, the decrease in inequality takes place across the distribution. The direction of evolution toward decreasing inequality in Figure 4 is qualitatively similar to that in Figure 2.

To some extent this decrease of inequality is due to the move of Chinese footprints into the middle of the distribution, which is represented by the longest straight segment visible in Figure 2, as pointed out by Lawrence et al. (2013) and documented for the global distribution of incomes by Lahoti et al. (2016), Milanovic (2016) and Nino-Zarazua et al. (2016). To illustrate this point, Figure 4 shows the trajectory of the 5$^{th}$ quantile of the Chinese carbon footprint distribution, which is the global ranking of the average carbon footprint of a Chinese person in the 40$^{th}$ to 50$^{th}$ percentile of the Chinese expenditure distribution. Between 1970 and 1995, its rank in the global distribution hardly changed. This group had a carbon footprint that was higher than roughly 56% of the global population. However, from 1995 to 2010 and even more in 2013, the position drastically improved to being higher than almost 70% of the global population, a ranking that is slightly higher than that quantile's position in the income distribution (Lakner and Milanovic, 2016). It is explained by the fact that China's average carbon footprint is estimated to be higher in relation to the global average footprint, than its average income is to that of the global average income.

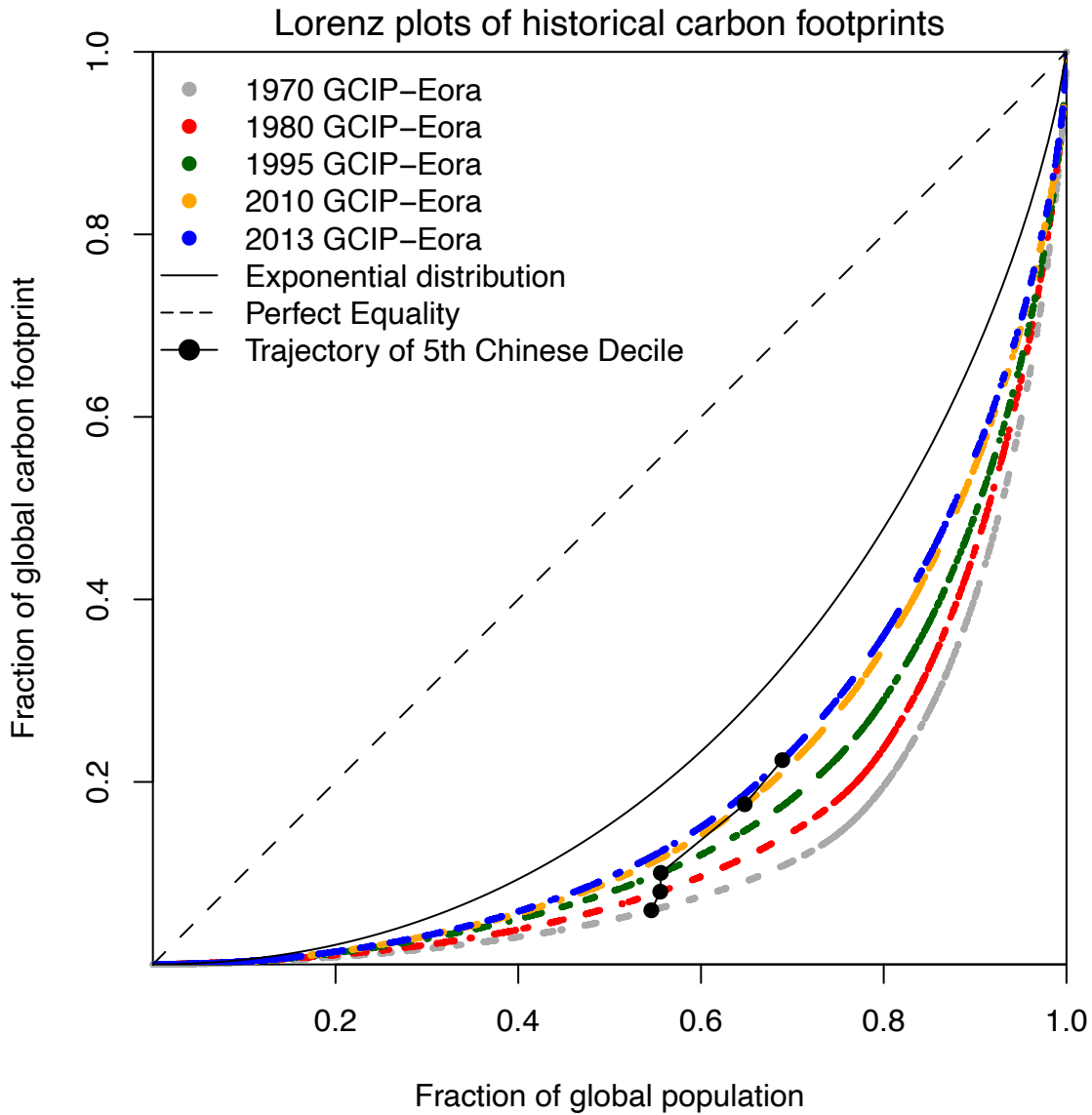

Figure 4: Lorenz plots of the Eora-GCIP data for selected years. Each dot represents an income quantile observation for a country. The solid line is the Lorenz curve implied by a theoretical exponential distribution. The line with black disks indicates how the Chinese 5th decile travels across the curve over time.

To get a broader, if still selective, overview over the movements within the Lorenz distribution, Figure 5 plots the trajectories from 1995 to 2013 of each of twelve quantiles in the Lorenz emissions ranking (i.e. their position on the x-axis in Figure 4) for nine countries that are large either in terms of emissions or population or both. The top row shows large high-income countries: France, Germany and the US. Their almost entire population continues to have per capita footprints that are higher than those of 80% of the global population over the time period, and even 90% for the US. However, it is also evident that the lowest two deciles have seen their position eroded. For instance, in France, the lowest decile dropped fifteen percentage points from 70% to 55%. As a result, the lowest carbon footprints, which correspond to the lowest incomes and expenditures, in the richest countries are now lower than those at the top of emerging economies.

This is visible in the middle row, with China, India and Indonesia. Most of the Chinese population has moved to well above a half of the global population in terms of carbon footprints by 2013. What is remarkable is the upward move of the entire Chinese distribution. A close visual inspection also shows that all quantiles of the Indian distribution have moved up, albeit just few percentage points. This change is from much lower levels, with only the top 10% appreciably higher than 50% of global footprints. In Indonesia, only the top 5% of emitters have moved up, whereas the rest of the population, at most, kept their rank. The third row shows that emerging and developing economies have an uneven development. While South Africa and Brazil have about a half of their population in the higher half of global emitters, their lower quantiles lost ground in the ranking. Nigeria's distribution has no overlap at all with rich country carbon footprints in the upper row, and the ranking of its top quantiles has not improved over 1995 levels. In other words, the overlap between rich and poor countries' emissions, with the exception of China, remains limited.

As pointed out already by Chakravarty et al. (2009), capping emissions of the richest means capping emissions of some individuals around the world. For instance, if the top 20% emitters had to reduce their emissions, this would, at least, affect the top percentile in all countries depicted, except Nigeria. But apart from this commonality, there is a large difference in who would be affected: the US would see 90% of its population affected, Germany 60%, and France 40%. China would see 30% affected, South Africa 20%, Brazil 5%, and India and Indonesia only 1%. While Grubler and Pachauri (2009) rightly caution about the accuracy of the detailed figures specifically for India's emissions and suggest there might be more high-footprint Indians, we can see that the 'global' high emitters in terms of footprints are still overwhelmingly concentrated in rich countries for our sample (analysis of other countries supports this conclusion). Among the countries depicted, only China sees a genuine upward movement in the position of its entire population.

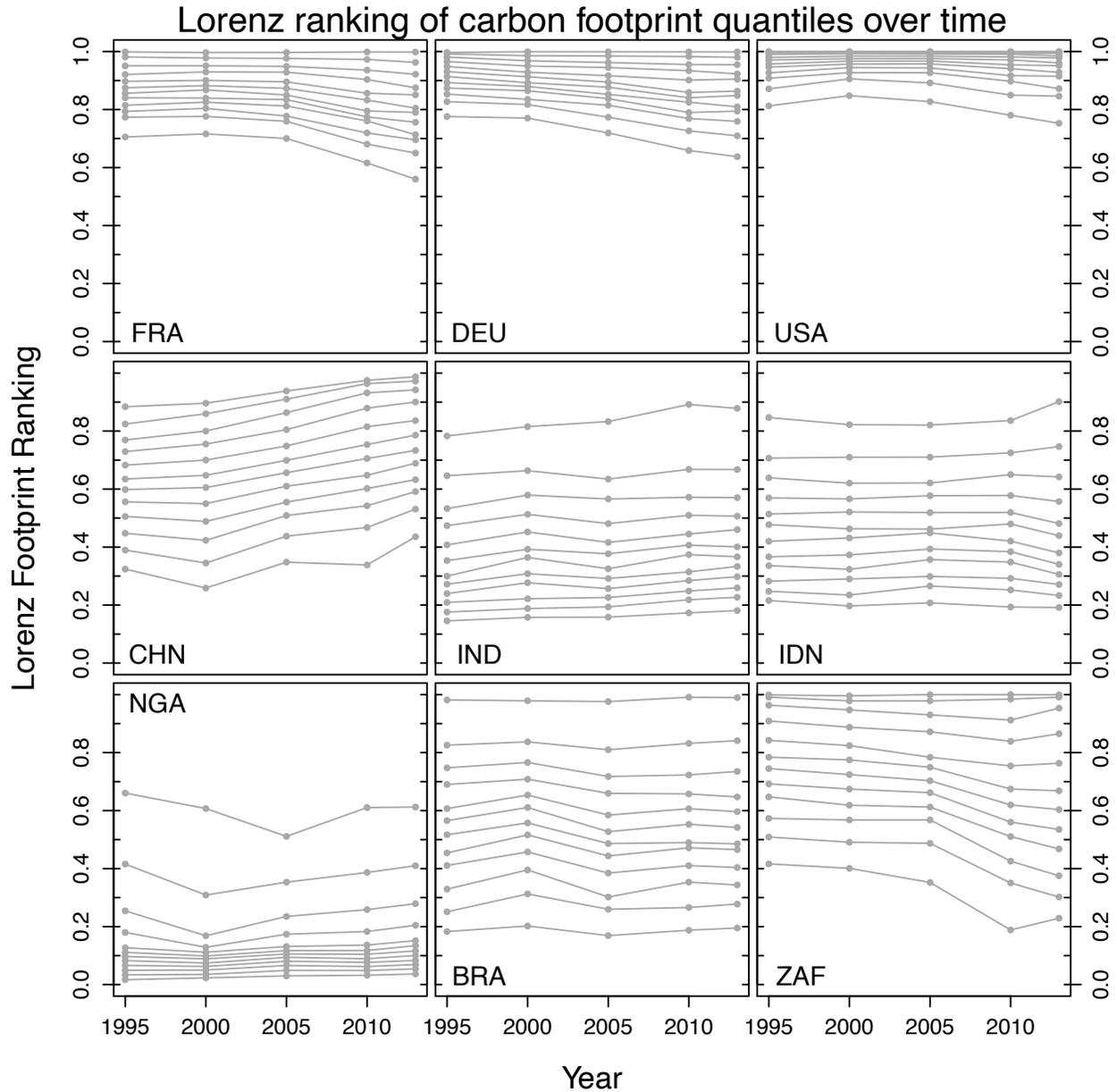

Figure 5: Trajectories of each quantile of selected countries through the Lorenz ranking. Each dot indicates for a quantile the share of global population that has a lower carbon footprint in that year. The bottom nine lines depict deciles, the top three the 90-95th quantile, 95-99th quantile, and top percentile.

The Lorenz trends shown in Figures 4 and 5 translate into Gini indicators decreasing from very high levels. Figure 6 depicts the Gini indices of the new GCIP-Eora estimates (black disks) and also compares them with existing ones (other symbols). The black disks show that the Gini coefficients for the global $CO_2$ footprint were consistently above 0.6. From a Gini of 0.73 in 1970, there was a downward trend that accelerated after 2000. Between 2000 and 2013, the global Gini fell by about 7 points, with the fastest decline in 2005-2010, which includes the Great Recession. To the extent that the series before 1995 are comparable (they exclude the Soviet

Union, which would add to the middle of the distribution and hence likely reduce inequality), inequality since 1970 fell about 13 Gini points, or roughly 3 points per decade.  The trend of decreasing Gini coefficient in Figure 6 is qualitatively similar to the trend in Figure 3, although the absolute levels in Figure 6 are substantially higher.

The 2010 GCIP-Eora estimate is remarkably close to the 2008 one by Chancel and Piketty, based on a different dataset, and is encouraging for the credibility of the results. The 2010 estimate by Hubacek et al. 2017, relying on a bottom up estimate of consumption baskets, is much lower. The difference between these estimates requires further examination. But part of the extent of inequality in the Hubacek data may be lost by the coarsely grained income quantiles.

Relative to any of these empirical estimates, both cap and floor scenarios are dramatically lower and imply halving the Gini estimate of global carbon footprint inequality. About 10 to 15 Gini points would have to be lost every decade after 2010 in order to get there, depending on the estimate. Settling on a middle value of 12 per decade, this decline would be unprecedented. Combining the two scenarios would require doubling this rate to achieve a Gini of about 0.07 (not shown in Figure 6). Even the fastest drop from 2005 to 2010 only saw the Gini decline by 3.5 points in five years, which extrapolates to 7 points per decade. This took place against the backdrop of the largest financial and subsequent economic crisis in rich countries since 1929, which eroded relative purchasing power and associated carbon footprints in their economies. Clearly, in comparison with historical levels and trajectories, scenarios that involve reductions in future carbon emissions are far away from anything seen in the recent past.

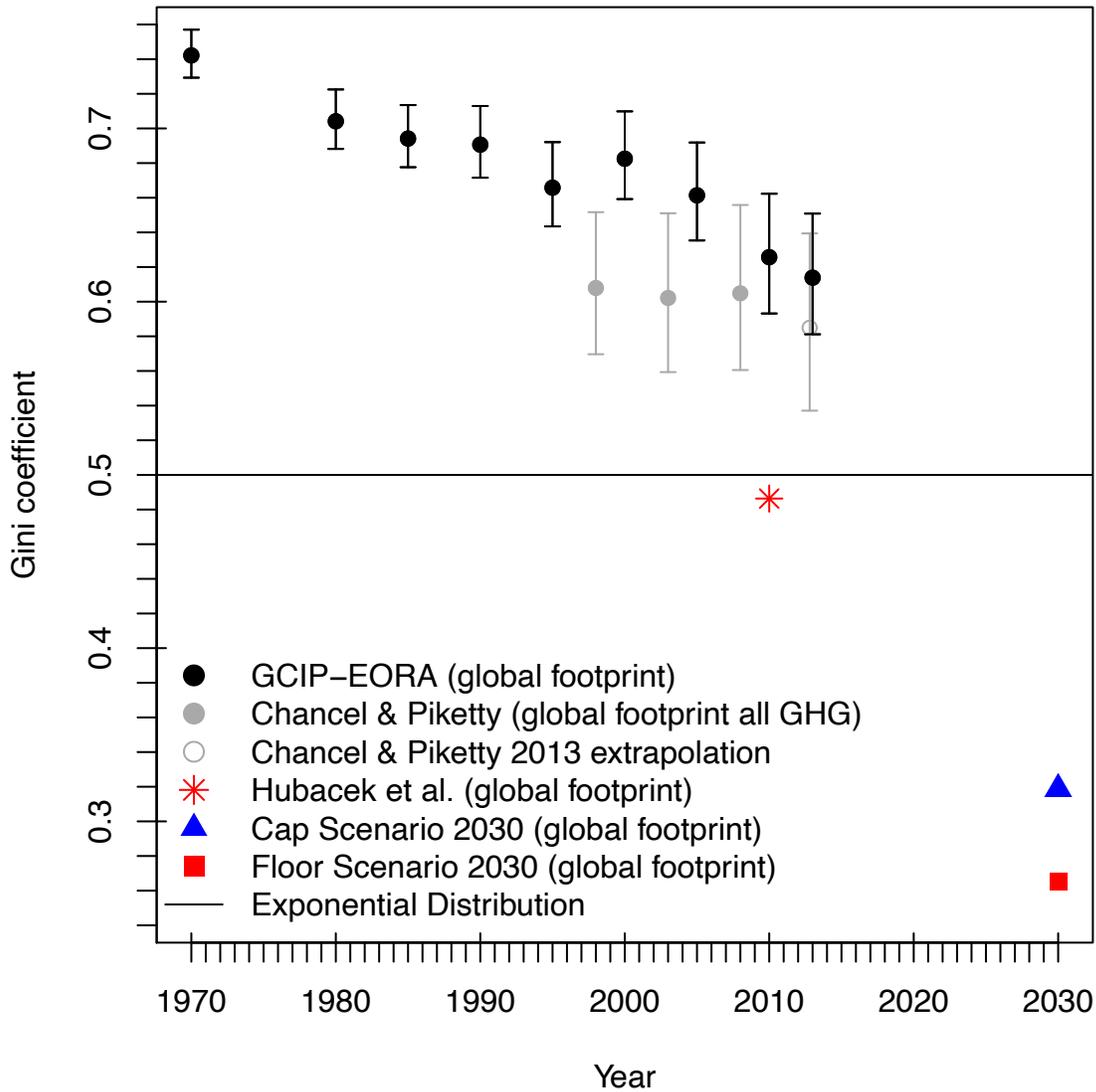

Figure 6 Historical evolution of the Gini coefficient for global carbon footprint. The new GCIP-Eora based estimate is shown by black disks for the income elasticity of carbon footprints equal 0.9 (lower and upper error bars for 0.7 and 1.1). The Chancel & Piketty estimate is shown by grey disks with the same elasticity error bars (the empty circle for 2013 is an extrapolation from 2008 data). The star, triangle, and square show the Gini for 2010 derived from Hubacek et al. (2017), and projected from the cap and floor scenarios for 2030. The horizontal line at Gini = 0.5 corresponds to a theoretical exponential distribution.

## 6. Discussion

How credible are the tremendous reductions in carbon footprint inequality envisioned by these ambitious scenarios? It appears that Rao and Min's (2018) assumption of a maximum 7 Gini points per decade decline in inequality is also about the most that has been historically observed at the global level for only half a decade and against the backdrop of an economic crisis predominantly in high carbon footprint countries. The previous section showed that policy reaching for cap or floor or both scenarios would seek to double this decline rate. What grounds would there be for more optimism about achieving ambitious carbon inequality reduction?

First of all, we point out that the historical data in Figure 6 show the decrease of global inequality in carbon footprints only up to 2013. However, the more recent historical data points in Figure 3 up to 2017 reveal saturation of the weighted international inequality in carbon emissions after 2013 at the level consistent with the maximal entropy principle. Given this saturation, predicted theoretically by Lawrence et al., (2013) and which is also applicable to global inequality, simply assuming the decreasing trend in carbon footprints inequality shown in Figure 6 after 2013 is unjustified.

One way to force down carbon footprint inequality would be to quickly reduce income inequality, as the two magnitudes are highly correlated.[4] The catch-up of China has recently led to a decrease in global income inequality (Lakner and Milanovic, 2016), and that is reflected in China's climb up the ranking in Figure 5 and the decline of carbon inequality in our Figures 4 and 6. But there is no inherent reason for this income inequality decline to persist, let along accelerate. Apart from China's move, the reduction in global measures of inequality comes mainly from the catch-up of non-Western elites with their rich-country counterparts. This decoupling of an elite is even more evident in increasing concentration of wealth among a small, global group of high net worth individuals (Davies et al., 2018, 2017; Piketty and Zucman, 2014). Reducing carbon inequality would require a fast break with current trends in income inequality.

There are, at least, two reasons to believe that high inequality in incomes is likely to persist. First, inequality has been and remains high between countries. Although there has long been hope for income convergence between rich countries and developing ones based on theoretical claims from neoclassical economic growth models about catch-up to the technology frontier, there is no evidence that developing countries as a group have converged with their high-income peers over the last 50 years, not even in terms of purchasing power parity (Johnson and Papageorgiou, 2020). This reflects the organization of the global economy: international trade and investment occur in the structured context of global capitalism. This organizational form generates and maintains international inequality, whether through overt colonialism predominant until the mid-20$^{th}$ century or a more indirect form of imperialism that produces unequal outcomes through its legal and regulatory architecture backed up by military superiority (Ghosh, 2019; Harvey, 2003; Kohli, 2020). Therefore, even though there is some catch up of top quantiles in developing countries, as Figure 5 shows, we should not expect that all of developing countries' emissions would catch up with high-income country emissions any time soon. Moreover, the resulting Lorenz curve after the catch-up for 2017 in Figure 2 is close to the exponential distribution, which maximizes entropy. Thus, one can argue that China's catch-up and the reduction of inequality from higher historical levels were implicitly driven by entropy maximization consistent with the constraints imposed by the current institutional structure of global capitalism. Now that the exponential distribution has been reached, the maximal entropy principle, which applies equally well to income inequality (Yakovenko, 2013), indicates preservation of the current level of inequality and the end of the previous decreasing trend.

Second, while international inequality in incomes has declined, inequality within countries has increased, certainly in richer countries (Alvaredo et al., 2018). This is often explained by regressive domestic economic policies and weakening of social safety nets combined with globalization, whereby global value chains put downward pressure on less-skilled workers' wages (or increase unemployment outright) in rich countries through competition with developing countries, while the incomes of managers at company headquarters grow faster (Aguiar de Medeiros and Trebat, 2017; Milberg and Winkler, 2013; Wood, 1995). And despite rhetoric to the contrary, the current US administration's policies have supported an entrenchment of these trends (Mayer and Phillips, 2019). Therefore, while inequality is responsive to institutional changes (Milanovic, 2016), reconfiguring the economic landscape so as to produce a double-digit Gini coefficient decline in income inequality in a matter of a decade appears highly unlikely. In this context, it is revealing that even the most optimistic "shared socioeconomic pathways" about future social trends, which are selected as the main scenarios for International Panel on Climate Change's forthcoming 6$^{th}$ assessment report, project only modest declines in income inequality in developing countries and rising inequality in high-income countries (Rao et al., 2019). In conclusion, to the extent that carbon remains coupled with income and expenditure, a fast decrease in carbon emissions would need to be forced against the structural tendencies towards high socio-economic inequality of the current economic system. The results in Figure 3 of a Gini coefficient converging to a (high) level of 0.5 also support this conclusion.

---

[4] In fact, the Gini graph for carbon emissions in Figure 3 is very similar to the Gini graph for weighted international income inequality (Concept 2) shown up to 2006 in Figure 1 of Milanovic (2012). This observation further confirms that global inequalities in income and carbon emissions are closely related.

This leaves open the question whether $CO_2$ footprints can be decoupled from income, which would enable both poverty eradication and capping emissions without interfering with the income distribution at the top. At the bottom of the footprint distribution, there is some hope of 'leap-frogging' dirty technologies, such as directly building electrified public transport infrastructure and skipping transport based on internal combustion engines. However, one problem with this approach is that it leaves unexplained just how the bottom billions of people could quickly improve their incomes. Historically this has happened via industrialization, which implies energy-intensive industrial processes (that are also carbon intensive, if energy is supplied from fossil fuels), such as steel and cement production (Semieniuk et al., 2019). The worry by development economists about "premature deindustrialization" suggests that there is as yet no other path even to modest prosperity (Rodrik, 2016).

At the top end of incomes, one could imagine consumption shifting to lower-carbon commodities in the luxury segment, possibly aided by strong incentives or regulation. For instance, richer people could source 100% of electricity in the household from renewables with back-up batteries, travel via luxury electric cars, commute by non-emitting air taxi, see e.g. Uber's Elevate project (Holden and Goel, 2016), and buy products that are certified produced with non-fossil energy. In such a scenario, the richest persons might eventually emit less per person than their middle-class counterparts, who cannot afford relatively expensive low-carbon products. This would make high-income inequality compatible with the cap proposal. However, current data on the consumption baskets of high net-worth individuals, which are not accurately reflected in the surveys we examined, but are best represented by the footprint of the top percentile, suggests the opposite. Lifestyles and occupations of the richest persons rely crucially on products that are hard to decarbonize, such as frequent long-distance flying – often in private jets (Gössling, 2019; Otto et al., 2019). Thus, the elasticity of the top emitters may be particularly high and the opposite scenario of rising carbon inequality at the top might obtain, as the number of millionaires and billionaires increases. Moreover, some products are low carbon only during usage, but not over their life cycle. For instance, current solar PV panel *manufacturing* methods emit significant amount of $CO_2$ (Miller et al., 2019). With a footprint approach, even if the manufacturing takes place abroad, this will still be attributed to the consumers of solar electricity. Therefore, direct measures regarding consumers would need to be complemented with decarbonizing in the industrial sector (Liu and van den Bergh, 2020). Finally, to the extent that emissions are saved through energy efficiency improvements, the monetary savings can lead to additional purchases of other more carbon intensive products, causing rebound effects (Feng et al., 2015; Thomas and Azevedo, 2013). Of course, even if highest-income lifestyles could be decarbonized, this dynamic would spell trouble of another kind. Since real-world emission reductions are likely to rely on a carbon price, the decarbonization of the rich's consumption baskets leads to climate change mitigation policy becoming even more regressive than it would be with current consumption patterns (Boyce, 2018). This leads right back to the problem of income inequality.

This ultimately points to the pivotal importance of decarbonizing the entire energy supply (not just electricity), which reduces everyone's carbon footprint, through the low-carbon energy embodied in final products, regardless of income distribution. Importantly, it would also remove the perverse effect of associating the eradication of poverty and reduction of inequality at the lower end with strong increases in carbon emissions. A low-carbon energy supply sidesteps the redistributive issues around demand-side decarbonization, because it decouples energy demand from carbon emissions at all income levels. The poorest households often do not yet have access to even the most basic energy services (such as electrification). For this segment of the population, there is therefore a particular opportunity to provision services from a low-carbon energy supply, and never install fossil energy generation in the first place (Brand-Correa et al., 2018; Pachauri, 2014). This decarbonization, and also poverty eradication, would work even as *relative* carbon inequality might remain high.

The foregoing analysis of the structural constraints on lowering carbon inequality has a close connection with the discussion of maximum entropy in Section 3. The maximum entropy approach predicts that emissions tend to a certain (rather unequal) distribution given constraints. In the case of territorial emissions, the weighted international distribution is well predicted by maximum entropy only with a constraint on total emissions to converge to an exponential one, at least as long as a large share of energy comes from tradable fossil sources. In the case of the footprint, our discussion suggests that other constraints might keep inequality at an even higher level. Institutional change would need to remove these constraints and erect others that keep inequality at a low level (Reddy, 2020). Our argument here is that given the evidence about current global economic structures and trajectories, it would be difficult to change these constraints in the manner and at the pace to attain low inequality sufficiently quickly.

# 7. Conclusions

This paper has quantified and analyzed international and global carbon emission and footprint inequality and contrasted historical trends with the projected inequality in scenarios that would cap and floor per capita footprints to achieve ambitious climate mitigation goals. The results show just how striking a departure from historical inequality trends may be required to achieve the twin goals of poverty eradication and climate change mitigation via that strategy. It is tempting to argue that the future will look different from the past, because ambitious mitigation and poverty reduction policies have not yet been tried. However, we also show based on our results and drawing on recent literature on global capitalism that it would be a herculean task for policy makers to overcome the structural forces of the current economic system that reinforces high socio-economic inequality within and between countries. We conclude that policy should focus strongly on decarbonizing the energy supply, so that climate change mitigation and poverty reduction would not require equality of carbon footprints.

An important result of our study is the empirical demonstration of the striking convergence of the distribution of weighted international inequality in territorial carbon emissions to an exponential form, implying stabilization of inequality at a 0.5 Gini coefficient. One of the authors predicted this convergence based on maximum entropy reasoning before it was evident in the available historical data (Lawrence et al., 2013). The subsequent confirmation shows that the predictions based on the combinatorial arguments about most likely distributions fare well with respect to carbon emissions. It also suggests that global carbon footprint inequality, which is estimated to be variously around 0.5 or higher, depending on the method of estimation, may similarly be bounded from below due to the prevailing constraints, adding more impetus to the demand for energy supply decarbonization.

This article has characterized the inequality in current global carbon footprint estimates and advanced arguments for why it may be relatively stable as long as the energy supply is mostly fossil fuels. Further research could attempt to collect the increasingly dense coverage of national income elasticity estimates of carbon footprints to improve the measurement of the extent and evolution of carbon and greenhouse gas footprints. It could also attempt to characterize constraints on inequality that arise from the current structure of the global economy and identify ways in which changes in structure, e.g. through deliberate policy, could overcome these constraints.

**Acknowledgements**
The authors have benefitted from discussions with Lopamudra Banerjee, Paulo dos Santos, Duncan Foley, Sanjay Reddy, Lance Taylor, Isabella Weber and from Scott Lawrence's updated Gini graph of weighted international $CO_2$ emissions. They are grateful to Klaus Hubacek and co-authors for sharing their carbon equivalent footprint data, and to Lucas Chancel and Thomas Piketty for making their data available online.